# DECIMETRIC EMISSION 500″ AWAY FROM A FLARING SITE: POSSIBLE SCENARIOS FROM *GMRT* SOLAR RADIO OBSERVATIONS.

Susanta Kumar Bisoi,[1] H. S. Sawant,[2] P. Janardhan,[3] Y. Yan,[1] L. Chen,[1]
Arun Kumar Awasthi,[4] Shweta Srivastava,[3] and G. Gao[5]

[1]*Key Laboratory of Solar Activity, National Astronomical Observatories,*
*Chinese Academy of Sciences, Beijing 100012, China.*
[2]*INPE - National Institute for Space Research, Sao Jose dos Campos, SP, Brazil*
[3]*Astronomy & Astrophysics Division, Physical Research Laboratory, Ahmedabad 380 009, India.*
[4]*Key Laboratory of Geospace Environment, Department of Geophysical and Planetary Sciences,*
*University of Science and Technology of China, Hefei 230026, China.*
[5]*Yunnan Observatories, Chinese Academy of Sciences, Kunming 650011, China.*

## ABSTRACT

We present a study of decimetric radio activity, using the first high time cadence (0.5 s) images from the *Giant Meterwave Radio Telescope* (*GMRT*) at 610 MHz, associated with *GOES* C1.4 and M1.0 class solar flares, and a coronal mass ejection (CME) onset that occurred on 20 June 2015. The high spatial resolution images from *GMRT* show a strong radio source during the C1.4 flare, located ∼500″ away from the flaring site with no corresponding bright footpoints or coronal features nearby. In contrast, however, strong radio sources are found near the flaring site during the M1.0 flare and around the CME onset time. Weak radio sources, located near the flaring site, are also found during the maximum of the C1.4 flare activity, which show a temporal association with metric type III bursts identified by the Solar Broadband Radio Spectrometer at Yunnan Astronomical Observatory. Based on a multi-wavelength analysis and magnetic potential field source surface extrapolations, we suggest that the source electrons of *GMRT* radio sources and metric type III bursts were originated from a common electron acceleration site. We also show that the strong *GMRT* radio source is generated by a coherent emission process and its apparent location far from the flaring site is possibly due to the wave-ducting effect.

Corresponding author: Susanta Kumar Bisoi
susanta@nao.cas.cn

## 1. INTRODUCTION

Decades of observations have shown that solar flares release energy as large as $10^{23}$–$10^{33}$ ergs via the process of magnetic reconnection, thereby heating the coronal plasma and accelerating particles (e.g. Priest & Forbes (2000); Aschwanden (2002); Benz (2008); Fletcher et al. (2011)). The accelerated electrons, while propagating in the corona, emit in hard X-ray (HXR) and also in radio wavelengths at different heights depending on the electron energy and coronal electron density. However, an intriguing question remains regarding the location of the electron acceleration. Several earlier studies have arrived at a consensus that the primary electron acceleration sites are located in the lower corona (Bastian et al. 1998; White et al. 2011), where the electron density $n_e$ is $10^9 - 10^{10} cm^{-3}$. The radio emissions detected from this region, if due to plasma radiation, usually occur in the decimeter wavelength range, $i.e. f \approx 500$ – 2000 MHz (Aschwanden 2002, 2005). It is, therefore, of particular interest in utilizing the decimetric flare emission to locate and diagnose the electron acceleration sites and in turn, the flare energy release region.

Solar radio emissions at decimetric and longer wavelengths are broadly categorised into five types: Type I, II, III, IV and V (Melrose 1980; Dulk 1985) and they usually show very high brightness temperature ($> 10^9 K$) due to their generation by a coherent emission mechanism (Dulk 1985). In a plasma radiation process, plasma Langmuir waves are excited by accelerated non-thermal electrons. They are converted to radio emissions at the fundamental ($f_p = 8980 \sqrt{n_e}$, where $f_p$ is in Hz and $n_e$ is in $cm^{-3}$) and/or harmonic ($2f_p$) frequencies (Wild et al. 1953; Wild & Smerd 1972; Melrose 1980; Dulk 1985; Bastian et al. 1998). Another possible coherent emission mechanism is the electron cyclotron maser (ECM) radiation process (Melrose & Dulk 1982). It has been proposed in order to explain narrow band (10–150 MHz), small source size (3 $\times 10^9$ cm), strong circular polarisation (100%) and high brightness temperature ($> 10^{12}$ K) of spike bursts usually accompanying with type IV bursts (Droege 1977; Slottje 1980). Spike bursts that accompany with type III bursts (Benz et al. 1982; Benz 1985), however, show low to intermediate circular polarisation. Generally, for ECM emission to be possible, it requires that $f_p$ must be less than $f_b$ ($= \frac{eB}{2\pi m_e}$, where $f_b$ and $B$ are, respectively, electron cyclotron frequency and magnetic field, and $e$ and $m_e$ are, respectively, charge and mass of electron) at the radio emission site (Melrose & Dulk 1982; Melrose 1991).

Using high time sampling spectrograph and/or imaging observations, decimetric flare emissions including fast drift type III bursts (Aschwanden & Benz (1995), Benz (2004, 2008) and references therein,Chen et al. (2013)), drifting pulsating structures (DPS; Kliem et al. (2000); Karlický et al. (2002); Karlický (2004); Karlický & Bárta (2007)), and short-duration narrow band spikes (Droege 1977; Benz et al. 1982; Benz 1985, 2004; Tan 2013; Chen et al. 2015) have been recognised and studied. The escape of radio waves at decimetric wavelengths from the flare site is, however, difficult due to the usually strong free-free absorption by the relatively dense plasma in the low corona (Aschwanden & Benz 1995; Aschwanden 2002). A more favourable condition for detecting radio sources at this wavelength range requires relatively weak free-free absorption along its way (Aschwanden 2002), which can be achieved when the energetic source electrons are injected into over-dense structures filled with hot coronal plasma or the radio waves are ducted along a density depleted magnetic flux tube outward. In the low beta corona, both open magnetic flux tubes and high arching closed loops can act as density depleted ducts (Duncan 1979). The ducting of streaming electrons generally causes a shift in the position of true radio source, which was first commonly seen in the single frequency records of Culgoora heliograph radio images (Stewart 1972; Sheridan et al. 1972; Duncan 1979). The true source is located at a lower coronal height where the radio wave is actually generated than that of the apparent source in the radio image. The trapped radio waves only leave the flux tube at the apparent source position when the wave frequency inside the tube is greater than the plasma frequency outside. Such displacements of the apparent radio source have also been seen in radio images from the *LOw Frequency ARray* (*LOFAR*) (Kontar et al. 2017; Mann et al. 2018) at metric wavelengths. Based on a study of fine structures of type III radio bursts using *LOFAR* observations, Kontar et al. (2017) suggested that the apparent source locations of fundamental and harmonic emission result from the scattering of radio waves by coronal density inhomogeneities. Recently, using *LOFAR* observations, Mann et al. (2018) reported type III radio bursts located higher up in the corona and showing non-uniform movements in their source positions. The authors suggested that the type III radio bursts were due to plasma radiation and the apparent shifts in their locations were caused by coronal propagation effects. In another study using *NRH* images at 450 MHz, Benz et al. (2002) reported low frequency decimetric spikes which were located about $20''$ – $200''$ away from the HXR and soft X-ray (SXR) sources.

In an earlier study of radio emissions at decimetric wavelength range using the *GMRT* at 610 MHz, Kundu et al. (2006) also reported radio sources located about $200''$ from the flaring active region. In the present paper, using the first high time cadence images at 610 MHz from the *GMRT*, we have studied the decimetric radio activity associated with C- and M-class flares and a coronal mass ejection (CME) onset, that occurred on 20 June 2015, and reported a decimetric radio source during the C-class flare to be located about $500''$ away from the flaring site. The time cadence of the present *GMRT* observations is 0.5 s, while the earlier reported *GMRT* observations had a maximum time cadence of 16.9 s (Subramanian et al. 2003), 2.11 s (Kundu et al. 2006) or 1 s (Mészárosová et al. 2013). The high time cadence images of *GMRT* 610 MHz during the C-class flare have also been used, in combination with the coronal images of extreme ultra-violet (EUV) and X-ray wavelengths, to investigate the genesis of radio emitting electrons of decimetric sources and their temporal association with the type III burst activity at metric wavelengths. The rest of the paper is organised as follows. Section 2 briefly discusses solar observations at the *GMRT*. In Section 3, observations and analyses of *GMRT* 610 MHz data on 20 June 2015 and the associated data at EUV, X-ray and metric wavelengths are discussed. Subsequently, in Section 4 we report spatiotemporal distribution, polarisation and magnetic configuration of *GMRT* 610 MHz sources, and the association with metric type III bursts. Finally, Section 5 and 6 interpret and conclude our results.

## 2. INSTRUMENT

### 2.1. *The Giant Meterwave Radio Telescope*

The *GMRT* (Swarup et al. 1991) is a radio interferometer and consists of thirty 45-m diameter antennas spread over a radius of 25 km. It is located in India, about 80 km from Pune. Fourteen antennas are in a compact configuration, randomly distributed in a central square area of 1 km while the remaining antennas are distributed along three arms in a Y-shape. The shortest baseline at *GMRT* is 100m while the longest baseline is 25 km. The long baselines of *GMRT* offer a high spatial resolution (upto $\sim 5''$ at 610 MHz), making it one of the most important interferometers operating in the low frequency range (150–1400 MHz). The GMRT actually observes at six distinct frequencies, viz. 150, 236, 327, 610, 1280, 1400 MHz. Though the GMRT is not a dedicated solar telescope, it has been used successfully for observing and studying the Sun (Subramanian et al. 2003; Kundu et al. 2006; Mercier et al. 2006; Mészárosová et al. 2013; Mercier et al. 2015). Snapshot solar images, with unprecedented resolution and high dynamic range, far better than any images previously obtained, have been produced by combining visibilities from the GMRT and the NRH (Mercier et al. 2006). This technique has been used by Mercier et al. (2015) to study the fine structures of solar noise storms using joint observations of the GMRT and the NRH and have achieved resolutions as small as 30″ at 327 MHz and 35″ at 236 MHz. Only few studies have been reported at GMRT 610 MHz. The first study (Kundu et al. 2006) of GMRT 610 MHz radio emission reported an arc-shaped radio source morphology of decimetric flare emissions describing their origin due to trapped radiating electrons at the magnetic loop tops. In another study of GMRT 610 MHz (Mészárosová et al. 2013), during a weak B8.9 flare, the authors reported very weak radio sources located in the fan structure of coronal magnetic field lines originating from a coronal magnetic null point.

The observational procedure during solar observations at GMRT have been described in Subramanian et al. (2003) and Mercier et al. (2006). For the present GMRT solar observation of 20 June 2015, the Sun was continuously observed first for one hour after which a nearby phase calibrator was observed for around 7 min. This cycle of Sun and phase calibrator observations was repeated during the entire allotted observation period. During each Sun scan, the antennas were positioned at the Sun center at the start of the scan and tracking was carried out at a linear rate in both Right Ascension (RA) and Declination (Dec). In subsequent scans, the antennas were again repositioned to the Sun center at the start of the scan and was tracked as before. In addition to the phase calibrator observations, a strong flux calibrator (such as 3C147) was also observed both at the start and the end of the observation.

## 3. OBSERVATION AND DATA ANALYSIS

Two GOES C-class (C2.3 and C1.4) and a GOES M-class (M1.0) solar flares were recorded on 20 June 2015 as seen from GOES15 SXR observations (Figure 1 (top)). The C2.3 flare was a short-duration flare, which started at 02:34:41 UT and ended at 02:44:00 UT while the C1.4 flare was a rather long-duration flare, that lasted for over 2 hrs. The C1.4 flare started at 03:14:41 UT and ended at 04:50:00 UT followed by an M1.0 flare that started at 05:57:21 UT and lasted for only 20 min. The onset time of the flares is determined by the time when the SXR flux is above the three sigma level of the pre-flare level. The vertical black dotted lines in the upper panel of Fig.1 demarcate the onset of the respective flares. Just after the M1.0 class flare, a CME was erupted at 07:36:00 UT as recorded in the C2 field of view of the Large Angle and Spectrometric Coronagraph (LASCO), which is indicated by a vertical red dotted line in the upper panel of Fig.1.

The GMRT radio observations of 20 June 2015 were recorded, from 03:47:39 – 11:09:16 UT, at 610 MHz with a time cadence of 0.5 s, covering the C1.4 and M1.0 flares, and the CME. The observed GMRT data were first properly edited and calibrated using both flux and phase calibrators. The calibrated data during the flare intervals were then processed using a standard clean procedure in Astronomical Image Processing System (AIPS), a standard radio astronomy software package, used primarily for data reduction of astronomical radio data sets. Generally, images so produced are registered properly, however, in order to make the image registration proper, we applied an additional calibra-tion. The calibration was carried out based on self-calibration solutions obtained at the peak of the burst during the M1.0 flare. The solutions were then applied to the radio data of both the C1.4 and the M1.0 flares and cleaned images were finally produced. Fig.1 (bottom) shows the *GMRT* 610 MHz light curve during the period 03:47:39 – 11:09:16 UT. The light curve was obtained from *GMRT* snapshot images made at 2s intervals. From Fig.1 (bottom), a strong flux enhancement during *GMRT* observations of the C1.4 flare period (indicated by a dotted red curve) is clearly evident in comparison to the flux enhancements during *GMRT* observations of the M1.0 flare and the CME (indicated by a dotted blue curve).

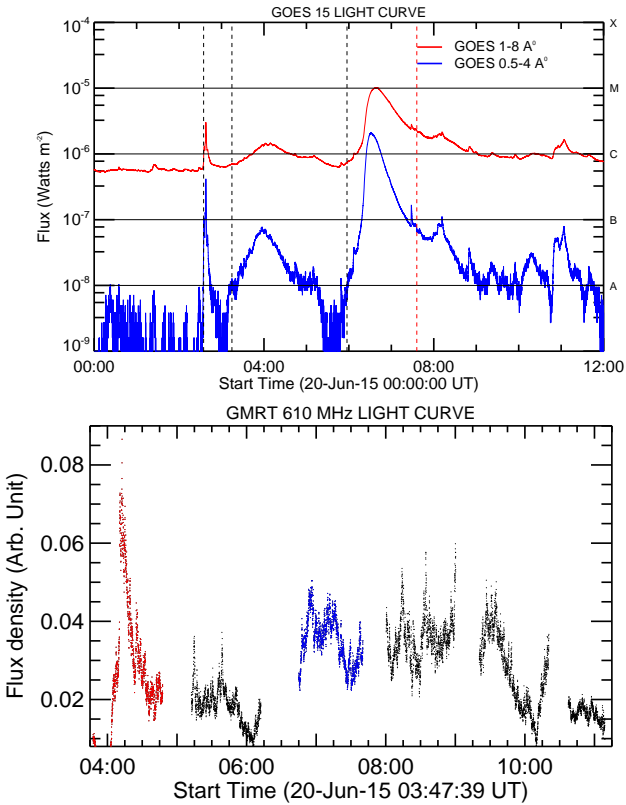

**Figure 1.** (top) The temporal profile of *GOES*15 X-ray observations on 20 June 2015. The vertical black dotted lines indicate the onset of *GOES* C2.3-, C1.4- and M1.0- class solar flares, respectively, at 02:34:41, 03:14:41 UT and 05:57:21 UT. The vertical red dotted line indicate the onset of the CME in LASCO/C2 field of view at 07:36 UT. (bottom) The *GMRT* 610 MHz light curve for the period 03:47:39–11:09:16 UT, covering the period of a C1.4 flare duration (indicated by a dotted red curve), and an M1.0 flare and a CME (indicated by a dotted blue curve).

For *GMRT* solar observations, usually, an attenuation of 30dB is used, so the absolute amplitudes of solar observations at *GMRT* are not properly obtained. Hence in order to estimate the absolute flux of the present *GMRT* 610 MHz observations, we compared the 610 MHz flux at *GMRT* to the solar flux level at 610 MHz obtained by the patrol telescopes of Radio Solar Telescope Network (*RSTN*) at Learmonth station. Both the flux levels at *GMRT* and *RSTN* for the C1.4 flare show a similar time structure, from 04:07:12 – 04:16:48 UT, around the radio flare maximum (figure not shown here). So we computed a scaling factor of 4000 to normalise the *GMRT* 610 MHz flux, with respect to the *RSTN* 610 MHz flux, covering the C1.4 flare observation period recorded at *GMRT*.

We used EUV observations from *AIA* instrument (Lemen et al. 2012) on board *SDO* (Pesnell et al. 2012) and X-ray observations from *Reuven Ramaty High Energy Solar Spectroscopic Imager* (*RHESSI*, Lin et al. (2002)). Full disk images from the selected AIA passband with high temporal (12 s) and high spatial (0.6″) resolution centred on 6 EUV wavelengths (94 Å, 131 Å, 171 Å, 193 Å, 211 Å, and 335 Å) were used along with the line-of-sight (LOS) magnetograms from the *Helioseismic Magnetic Imager* (*HMI*; Schou et al. (2012)) having a time cadence of 45 s and a spatial resolution of 0.6″. *RHESSI* images for different energy bands, spanning the range 6–50 keV, were reconstructed using the CLEAN method (Hurford et al. 2002) with a spatial resolution of 2″.3. The integration time for the *RHESSI* images is 1 min when the *RHESSI* count rate is low and 20s when the count rate is high.

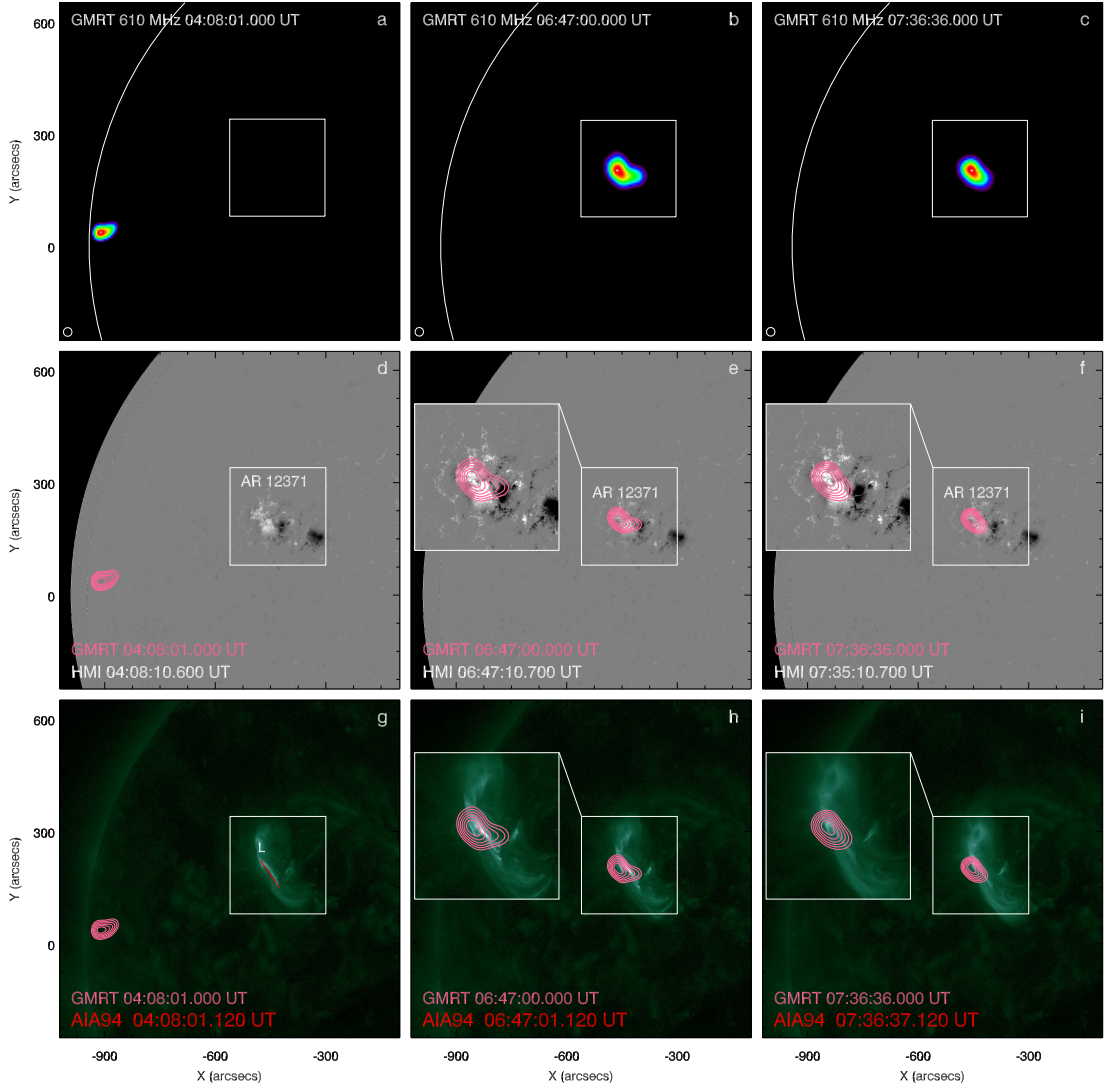

**Figure 2.** The columns, from left to right, show images, respectively, at the peak times of C1.4 and M1.0 flares, and at the onset of the CME, on 20 June 2015. The top row shows the *GMRT* 610 MHz snapshot maps, while the middle and bottom rows show their contour images (in pink) overlaid, respectively, on *HMI* photospheric and *AIA* 94 Å EUV images.

All EUV, X-ray and radio images, during the C1.4 and M1.0 flare observation periods, were derotated to their respective radio flare peak at *GMRT* 610 MHz to account for the effects of the solar differential rotation.

In addition, the radio dynamic spectra (the solar flux distributions in frequency-time plane) at low frequencies were obtained from the Solar Broadband Radio Spectrometer at *Yunnan Astronomical Observatory* (SBRS/*YNAO*, Gao et al. (2014)). The *YNAO* dynamic spectra used comprises of metric spectrograph observations in the frequency range, 70–700 MHz, in both left and right circular polarisations. The *YNAO* data have a spectral resolution of 0.2 MHz and a temporal resolution of 80 ms. However, due to severe interference in channels below 130 MHz and channels above 300 MHz, we presented only the interference free metric spectra in the frequency range 130–300 MHz.

## 4. RESULTS

### 4.1. *Spatial distribution of GMRT 610 MHz radio sources*

Snapshot images at *GMRT* 610 MHz at the flare maximum of the C1.4 and M1.0 flares and at the onset of the CME are shown in Figure 2 (top). The images have a clean beam of the order 12″, indicated in the bottom left corner in the top row panels of Fig.2. The signal-to-noise ratio at the peak of the burst during the C1.4 and M1.0 flares was 306 and 377, respectively, and the background noise level during the low flux periods was $3.5 \times 10^5$K. The middle row of Fig.2 shows *GMRT* 610 MHz contours in pink overplotted on background *SDO/HMI* images. The contour levels are drawn at 40%, 50%, 60%, 70%, 80% and 90% of the maximum brightness temperature in each image. The *SDO/AIA* 94 Å images overplotted by *GMRT* 610 MHz contours, in pink, are shown in the bottom row of Fig.2.

A radio source, located above the flaring active region, AR 12371, at the time of the M1.0 flare maximum and the CME onset, is seen from Fig.2b,c. In contrast, a radio source, at the C1.4 flare maximum, is seen located near the southeast limb, about 500″ or 365,000 km away from the flaring active region. No other active region are seen nearby the location of this radio source from *SDO/HMI* images (middle row) or any corresponding coronal features such as coronal loops from *SDO/AIA* 94 Å images (bottom row). This raises the question as to whether the observed limb radio source during the C1.4 flare is real. Our analysis, which shows radio sources located near the flaring active region and co-spatial with the hot coronal loops (middle and bottom rows) during the M1.0 flare and CME, suggests that the image registration is proper. Thus, we conclude that the radio source observed during the C1.4 flare is real.

It is found that the electron plasma density in the corona usually varies in the range 1.2–4.6 $\times 10^9$ cm$^{-3}$ if the radiations are due to fundamental and harmonic plasma radiations at a frequency of 610 MHz. The electron plasma density above active regions generally falls in this range and so one expects that the *GMRT* 610 MHz radio source would be located near the active region, which is actually not the case for the *GMRT* 610 MHz source observed during the period of C1.4 flare. Hence, it is interesting and unique, and henceforth we mainly focus on the nature and the origin of this *GMRT* radio source, and the study of *GMRT* radio observations associated with only the C1.4 flare period. A detailed analysis of *GMRT* 610 MHz observations of the M1.0 flare and CME evolutions will be described in a subsequent paper.

### 4.2. *Spatial distributions of GMRT 610 MHz sources during the C1.4 flare*

Figure 3a shows a *GMRT* 610 MHz map, during the radio flare maximum, at 04:11:05 UT (an animation of Fig.3a is available in the online version of this journal, which shows *GMRT* 610 MHz snapshots, at 2s interval, from 04:02:45–04:11:59 UT, covering the rise to peak phase of the 610 MHz radio flare). As discussed earlier, a radio source, labelled as LS, is located to the south-east limb. The source is first appeared at 04:02:45 UT (see online version of Fig.3a) having low source flux density. With time the source flux density increases and a strong flux density is observed from 04:10:23 UT. Just at the same time, weak sources, located near the flaring active region AR12371, are also observed (see online version of Fig.3a). The weak sources are well resolved by the *GMRT* clean beam and shown in the insert of Fig.3a, labelled as S1, S2 and S3. The flux density of the weak sources are above the background noise level and hence we consider them as real.

We estimated the sizes of *GMRT* sources by using a two dimensional Gaussian fit to the full width at half maximum of source emission regions and found sources S1 and S2 have appar-ent angular sizes of $\sim 20''$ or source dimensions of $1.4 \times 10^9$ cm. Similarly, source S3 has an apparent angular size of $\sim 10''$ or $7.2 \times 10^8$ cm. Source LS is found to have an apparent angular sizes of $33'' \times 20''$ or source dimensions of $2.3 \times 10^9 \times 1.4 \times 10^9$ cm. We computed the brightness temperature of all GMRT radio sources, following the method as described in Subramanian et al. (2007), which is indicated by colour bars shown in Fig.3a. All the *GMRT* sources have high brightness temperature $> 10^{11}$ K with source LS having a brightness temperature of $\sim 1.4 \times 10^{12}$ K, about an order of magnitude higher than that of sources S1, S2 and S3.

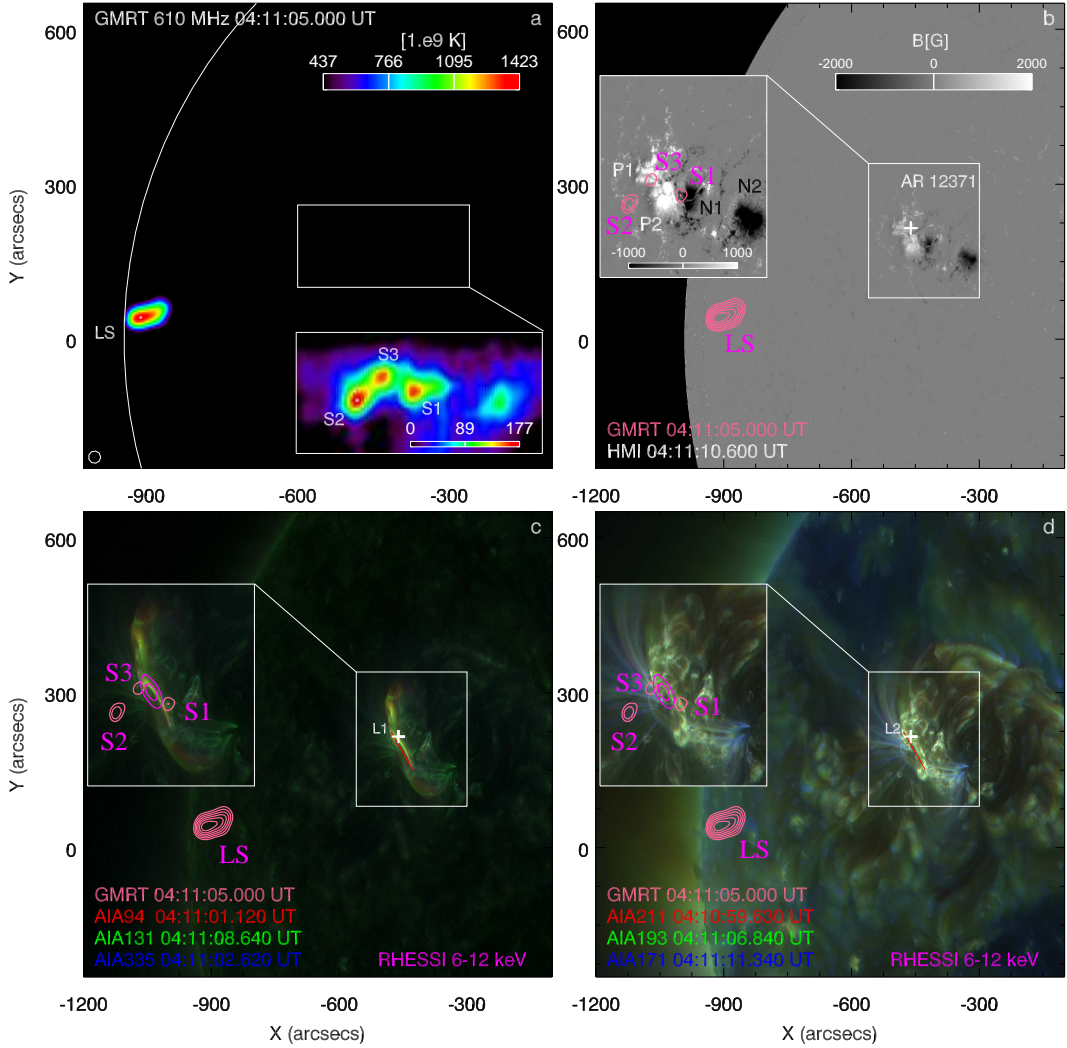

**Figure 3.** a) A *GMRT* 610 MHz snapshot map at the radio flare peak depicting a radio source at the south-east limb and three active region sources near the flaring location shown in the insert map. b) The overplotted *GMRT* contour images on an *HMI* magnetogram which shows a complex active region AR 12371 with a pair of positive (white) and negative (black) polarity patches. c) Hot flare loops, from the composite *AIA* passbands of 94 Å, 131Å and 335 Å, and d) long overlying and interconnected loops, from the composite *AIA* passbands of 211 Å, 193 Å and 171 Å, are observable near the flaring location. Also, *GMRT* (pink) and *RHESSI* 6–12 keV (violet) contours are overplotted on AIA hot and cold channel images. The straight curves, labelled as L1 and L2, are paths along which the time distance maps of Fig.5 are constructed. An animation of panels a and b of the figure is available in the online version of this journal. The animation covers the GMRT 610 MHz images from 04:02:45 to 04:11:59 UT while the static Figure displays this data at 04:11:05 UT.

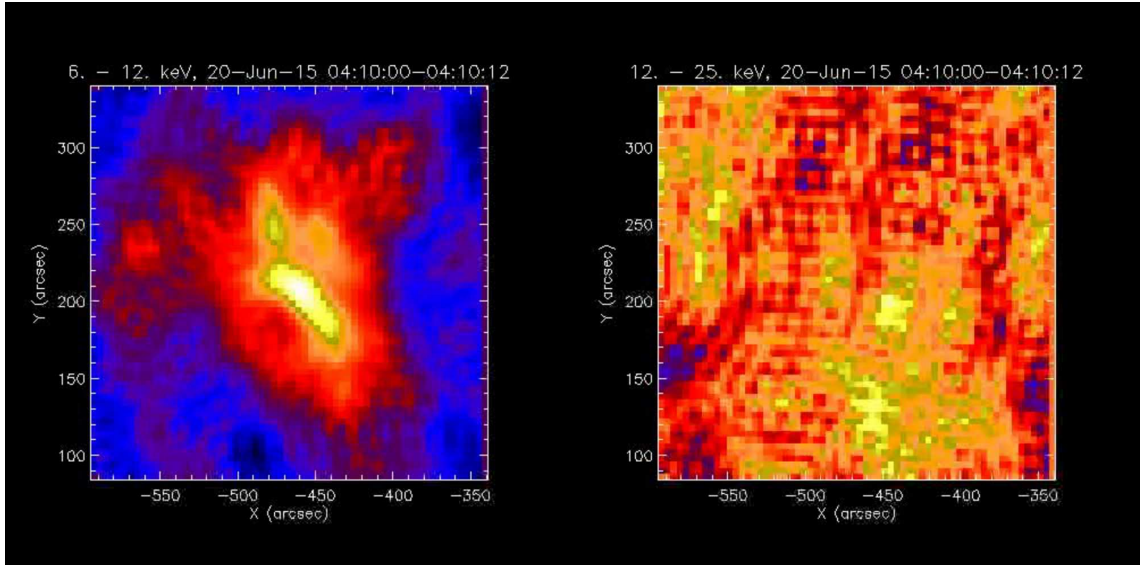

**Figure 4.** *RHESSI* clean images in a) 6–12 and b) 12–25 keV energy bands at 04:10 UT. The online animated version of this figure covers the period from 04:10–04:12 UT.

The overplotted *GMRT* source contours on LOS photospheric *HMI* magnetogram are shown in Fig.3b (An animation of Fig.3b is available in the online version of this journal, showing the overlaid *GMRT* source contours on *HMI* images from 04:02:45–04:11:59 UT). The C1.4 solar flare occurred in active region AR12371, the location of which is indicated by a white cross in Fig.3b,c,d. The active region AR12371 is shown in the insert of Fig.3b. It is a complex active region having pairs of sunspots of positive polarity, indicated by P1 and P2, and negative polarity, indicated by N1 and N2. From the overlaid *GMRT* source contours on HMI images, it can be seen that the source S1 is located above N1 while the source S3 is located above P1. The source S2 is displaced southward, about $40''$ to the east of the flare location with no corresponding magnetic features in the magnetogram. The source LS, however, has shown even further displacement of about $500''$.

The *GMRT* source contours overplotted on coronal images from *SDO/AIA* are shown in Fig.3c and d. Fig.3c shows a composite image constructed from *AIA* hot passbands of 94Å, 131Å, and 335Å, while Fig.3d shows a composite image constructed from *AIA* cold passbands of 211Å, 193Å, and 171Å. A partially formed sigmoid is visible from AIA hot channel image. The sigmoid is visible in the *AIA* cold channel image after the C1.4 flare maximum. At the northern end of the sigmoid, we can see hot coronal loops present near the flare location.

We prepared *RHESSI* clean images in 6–12 and 12–25 keV energy bands corresponding to the period 04:10–04:12 UT, when the weak

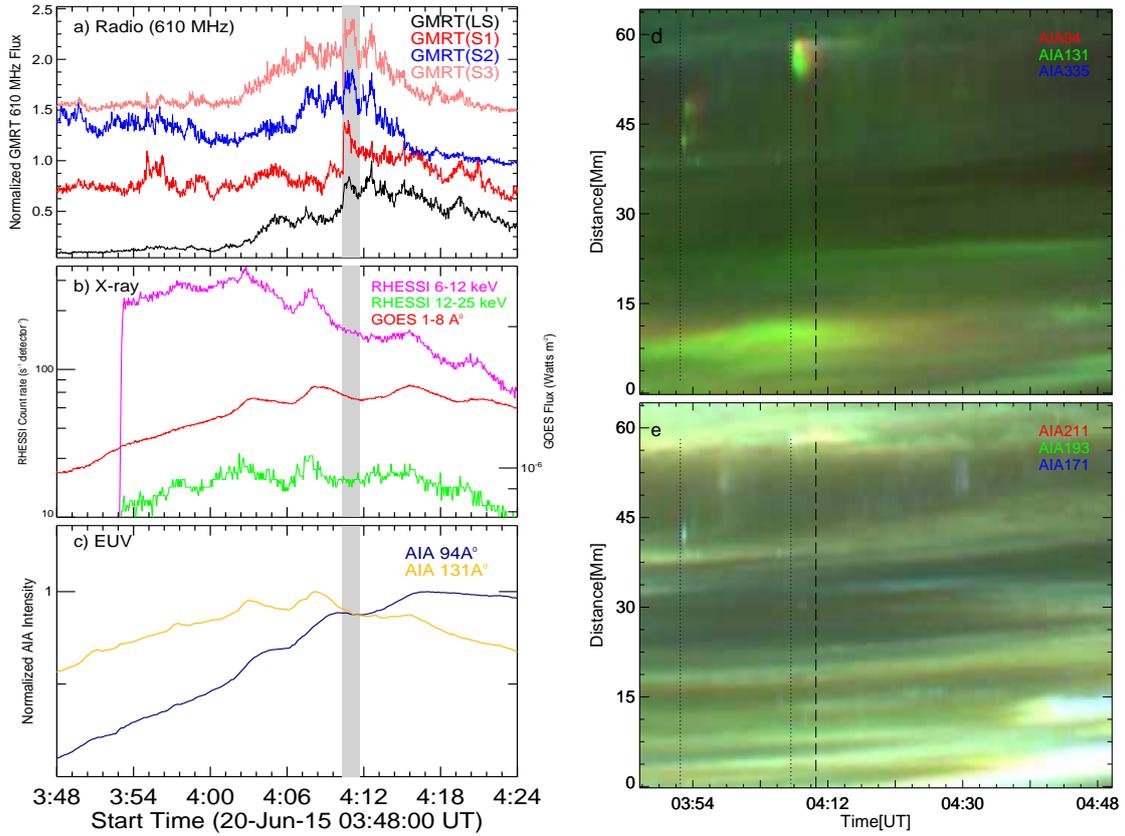

**Figure 5.** a) Normalised radio fluxes of *GMRT* 610 MHz sources LS, S1, S2 and S3 b) *GOES* flux in 1–8 Å channel and *RHESSI* count rates in energy bands, 6–12 and 12–25 keV. c) Normalised EUV intensities in *AIA* hot passbands of 94 Å and 131 Å. The radio flare maximum, corresponding to our time of interest, is indicated by a vertical grey bar in all panels from a-c. d-e) Distance-time maps of *AIA* hot and cold channel maps showing flare brightenings, marked by the vertical dotted lines, respectively, at 03:52:20 UT and between 04:07:25 UT and 04:10:25 UT.

*GMRT* sources have been identified near the flaring active region. An animation of the *RHESSI* clean images in 6–12 and 12–25 keV energy bands is available in the online version of Fig.4. From *RHESSI* 6–12 keV animation, a SXR source is clearly visible during the entire period. However, no consistent source structure is seen in *RHESSI* 12–25 keV images during the same period. Contour of high resolution (∼4s) *RHESSI* 6-12 keV image (in violet) made at 04:11:05 UT is also overlaid in insert maps of *AIA* hot and cold channel images, which shows a clear SXR source. But no HXR sources are seen in the high resolution *RHESSI* 12–25 keV image made at 04:11:05 UT and hence no HXR footpoints are observed (within the *RHESSI* sensitivity). It is seen from Fig.3c,d that the radio sources S1 and S3 are nearly co-spatial with the SXR source. Also, as seen from Fig.3b, the sources S1 and S3 lie on enhanced magnetic field regions. This suggests that the sources S1 and S3 could be radio sources located near the flaring active region. The source S2 and LS, on the other hand, are located far from the SXR source. Also, the sources S2 and LS when projected on the *HMI* magnetogram (Fig.3b), we find no magnetic field enhanced regions indicating they could not be located near the flaring active region. They presumably could be located on tall long loops (Kundu et al. 2006) and at relatively higher heights in comparison to the sources S1 and S3.

### 4.3. *Temporal evolution of GMRT 610 MHz sources during the C1.4 flare*

The temporal evolution of radio fluxes of all the *GMRT* 610 MHz sources, during the GOES C1.4 flare, are shown in panel a of Figure 5. The radio lightcurves of all the sources were obtained from *GMRT* solar maps by averaging solar fluxes from the surrounding region of the respective sources. The radio light curve of source LS (solid black curve in panel a of Fig.5) shows a flux level enhancement starting from 04:02:45 UT, which continues to increase until around 04:10:23 UT just when an abrupt enhancement in its radio flux level is noticed. Subsequently, its flux level shows a first peak, during the radio flare maximum, indicated by a vertical grey bar in the left hand panels of Fig.5. A second peak (maximum) is seen at 04:12:37 UT. Source S1 (solid red curve in panel a of Fig.5) shows a similar variation in its flux level, like source LS, with an abrupt enhancement in its flux shortly after 04:10:25 UT and afterwards peaking at 04:10:47 UT. The fluxes for sources S2 (solid blue curve in panel a of Fig.5) and S3 (solid light red curve in panel a of Fig.5) are also enhanced during this period peaking at 04:11:08 UT and 04:10:33 UT, respectively. The fluxes of sources S2 and S3, after the radio flare maximum, declines within few min while the fluxes of sources S1 and LS gradually declines showing a steady emission in the flare decline phase. It is noticed that a burst peak is present in case of all the *GMRT* sources during the 610 MHz flare maximum, although they showed different peak times. This suggests that the source electrons of all the *GMRT* sources could have access to a common electron acceleration site. The different peak flux times of all the sources though indicate that they could be located in different loop systems, which is apparent from their different spatial locations as seen in Fig.3.

The enhanced X-ray and EUV emissions are evident during the 610 MHz flare as noticed from *GOES15*, *RHESSI*, and *SDO/AIA* (94Å and 131Å passbands) light curves (Fig.5b, c). There is however, no one-to-one temporal correspondence between 610 MHz and X-ray fluxes (*GOES* and *RHESSI* 6–12 keV) during the radio flare maximum. This is commonly noticed in case of decimetric frequency bursts (Benz et al. 1983). It is observed that a *RHESSI* 6–12 keV X-ray burst occurred about 2 min prior to the radio flare peak. Such delay in the radio to SXR emission has been noticed (Aschwanden & Benz 1995) for decimetric type III bursts which peaked few min after the soft X-ray burst. Very low counts in *RHESSI* 12–25 keV energy band and hardly any counts in *RHESSI* 25–50 keV (not shown here) energy band indicate that there is very weak/no nonthermal X-ray emission (within the sensitivity of *RHESSI* instrument) around the radio flare maximum.

The distance-time plots were constructed by extracting the EUV intensity from *AIA* hot and cold channel images along the straight curve paths labelled L1 and L2 in Fig.3c and d, respectively. A first flare loop brightening is observable at 03:50:20 UT from the distance-time plots of *AIA* hot and cold channel images (Fig.5d,e), while the flare loops brightened later between 04:07:25 and 04:10:25 UT as seen only in hot channel images (Fig.5d), corresponding to the 610 MHz radio flare maximum period when the burst peak in all the *GMRT* sources has been observed.

In order to determine the thermal and nonthermal characteristics of the flare plasma during the radio flare maximum, we analysed the *RHESSI* X-ray spectra. We prepared a time series of background subtracted X-ray spectra during 04:10–04:12 UT at a time cadence of 12 sec. The background was determined by averaging the observations during 01:18:45 - 01:20:25 UT, corresponding to the closest-in-time quiet period. From the *RHESSI* X-ray count rates (figure not shown here), we find that the *RHESSI* count rates have been strongly affected by the particle emission, particularly in the energy bands > 50 keV. Thus, in order to indemnify the effect of high-energy particles, the time profile of the background was set to the 'high-E profile' option provided in the *Object Spectral Energy Executive* (*OSPEX*) package within the SolarSoft distribution. This option makes use of the intensity profile recorded in the highest energy-band, which in our case is 100-300 keV, to compute a time-dependent ra-tio. This ratio is subsequently applied on the background value, estimated by averaging the flux recorded during the quiet period (01:18:45 - 01:20:25 UT) in all the energy bands, enabling us to derive a time-dependent background evolution which then serves as an input to further spectral fitting. Background subtracted X-ray spectrum is evidently above the background level only up to the energy < 12 keV. We thus fitted the background subtracted spectra for each time interval, with an iso-thermal photon model available within OSPEX. The fitted spectra are shown in Fig.6 for a few selected time intervals (panels a-c), while the temperature and emission measure of flare plasma (panels d-e), are found to be varying in the range 7.7-9.1 MK and 0.9-3 $\times 10^{48} cm^{-3}$, respectively. From Fig.6, it is seen that the X-ray photon flux beyond 10 keV approximates to the background flux and hence the non-thermal characteristics of the flare plasma could not be determined with certainty from the spectral analysis.

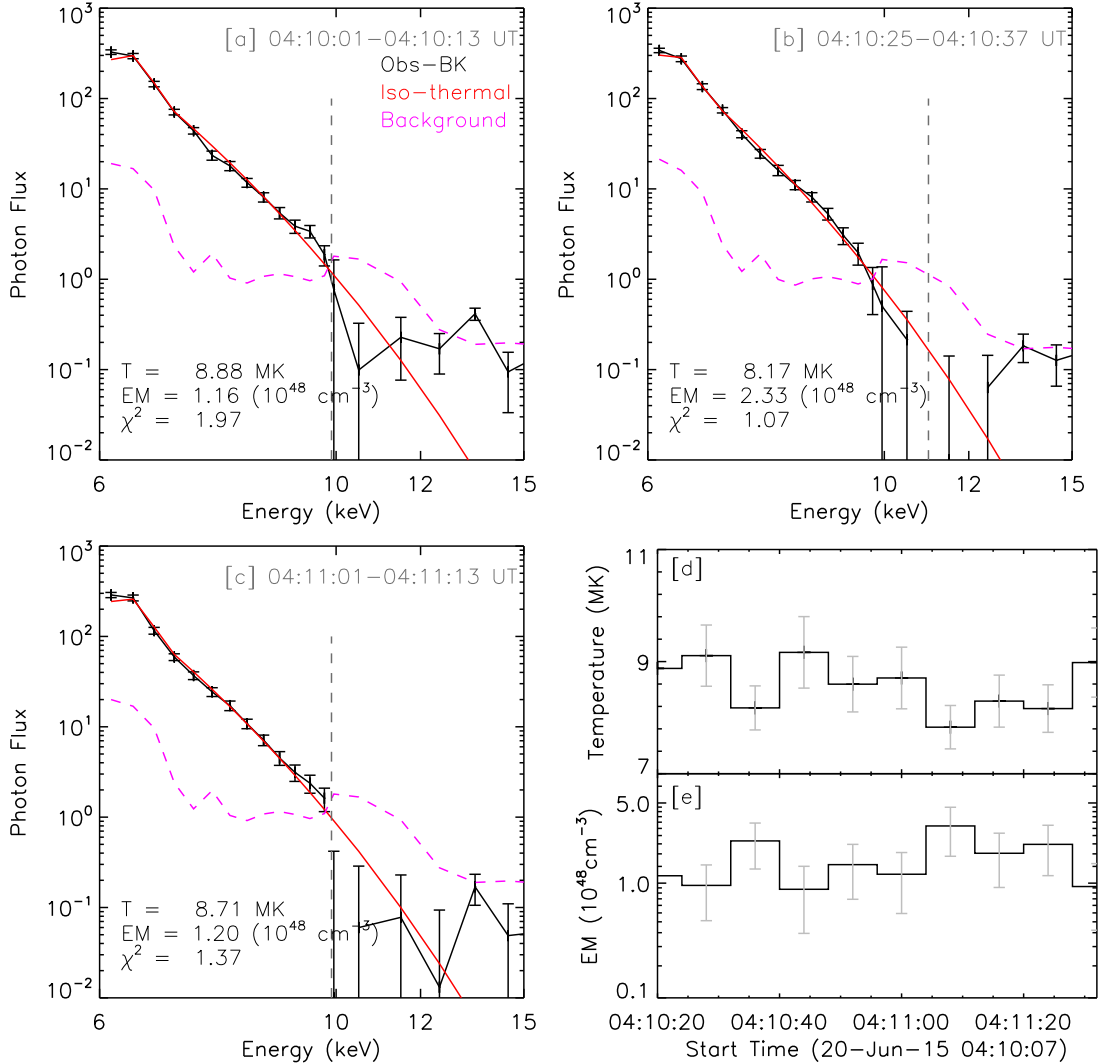

**Figure 6.** a-c) *RHESSI* background-subtracted X-ray spectra (black) above 6 keV shown for a few selected time intervals during 04:10–04:12 UT, corresponding to the radio flare maximum. Shown, in each panel, the dotted pink curve which is the background of spectrum, while the forward-fit to the background-subtracted spectrum, using an iso-thermal model, is shown by a solid red curve. In all the panels, the vertical line indicates the high energy end upto which the spectra are deemed to be fitted and also indicated are the values of temperature, emission measure and goodness of the fit (measured by $\chi^2$). d-e) Show temperature and emission measure of the flare emission, determined from the fitted X-ray spectra, during the time interval 04:10–04:12 UT.

### 4.4. *Polarisations of GMRT 610 MHz sources*

Fig.7 shows a sequence of *SDO/HMI* images overlaid by *GMRT* 610 MHz radio source contours in pink. The top row shows images in right circular polarisation (RCP), while the bottom row shows images in left circular polarisation (LCP). The images from left to right show radio sources, respectively, at the onset of the 610 MHz flare at 04:10:25 UT, at the 610 MHz flare peak at 04:11:05 UT, and after the 610 MHz flare peak at 04:11:45 UT. Source LS has been found in both LCP and RCP images at all times having relatively strong emission in LCP images. Thus, source LS is weakly left circular polarised throughout, with a degree of circular polarisation DCP = $(I_R-I_L)/(I_R+I_L)$ of ∼10%, where $I_R$ and $I_L$ are intensities of RCP and LCP, respectively. Source S1 is though seen in both LCP and RCP pre- and post-burst images, it is however, found only in RCP images during the burst, located above the region of negative magnetic polarity. Thus, source S1 is strongly right circular polarised having a DCP of ∼50%.

Sources S2 and S3, on the other hand, have been observed only in the burst images. Source S3 is seen only in LCP burst images with a DCP of ∼60%, located above the regions of positive magnetic polarity. While source S2 is found in both RCP and LCP burst images with a DCP of ∼10% and having relatively strong emission in LCP images. Hence, it is weak left circularly polarised. The strong polarisation of sources S1 and S3 indicates they are likely associated with fundamental radiations, while the weak polarisation of sources LS and S2 indicates they are possibly due to harmonic radiations.

### 4.5. *Magnetic field configuration of GMRT 610 MHz sources*

In order to understand the nature and origin of *GMRT* 610 MHz radio sources, their spatial image locations are compared with the coronal magnetic fields. The coronal magnetic fields were computed using photospheric magnetic fields from a *HMI* magnetogram at 04:11:05 UT as input to a potential field source surface extrapolation (PFSS) (Schrijver & De Rosa 2003). The computed extrapolated magnetic fields are overplotted on background *HMI* and *AIA* 94 Å images as shown in Fig.8. For clarity, only few coronal magnetic field lines are plotted. The magnetic field lines fan out from the magnetic active region AR12371 into the corona forming either the closed loops (indicated by white lines) or the open field lines (indicated by green lines). The smaller loops are mostly located near the active region while the larger ones spread out in all directions with one of their ends rooted in the active region. The contours of *GMRT* radio sources, in blue, are overplotted on the background magnetogram and EUV images.

It is noticed that source S2 is located near the loop top region of a loop, labelled as L1 in Fig.8 as solid pink curves, one of whose ends is at the positive polarity region P1 of active region AR12371, while the other end is at the negative polarity region N2 of active region AR12371. While source LS is located near the loop top region of a very long loop, labelled as L2 in Fig.8 as solid pink curves, with one of its ends rooted in active region AR12371, near the positive polarity region P1, and the other end located far away somewhere near the south-east limb. Although there could be other loops located along the line-of-sight of *GMRT* radio sources S2 and LS, we conclude that the two sources are possibly from the looptop region. By comparing the extrapolation results, we conclude that sources S1 and S3 are possibly from the legs of the flaring loop. This is implied from their positions near the ends of the extrapolated loop, shown by a pink curve, that lies on hot flare loops as seen in *AIA* 94 Å image of Fig.8 (the right panel). The extrapolated loop, intersecting source S1 and S3, is originated too in the positive polarity region P1. Overall, the magnetic field configurations associated with all the four *GMRT* sources possibly have a common origin. It is further noticed that the open field lines anchored near the positive polarity region P1 appear to be strongly divergent at larger coronal heights.

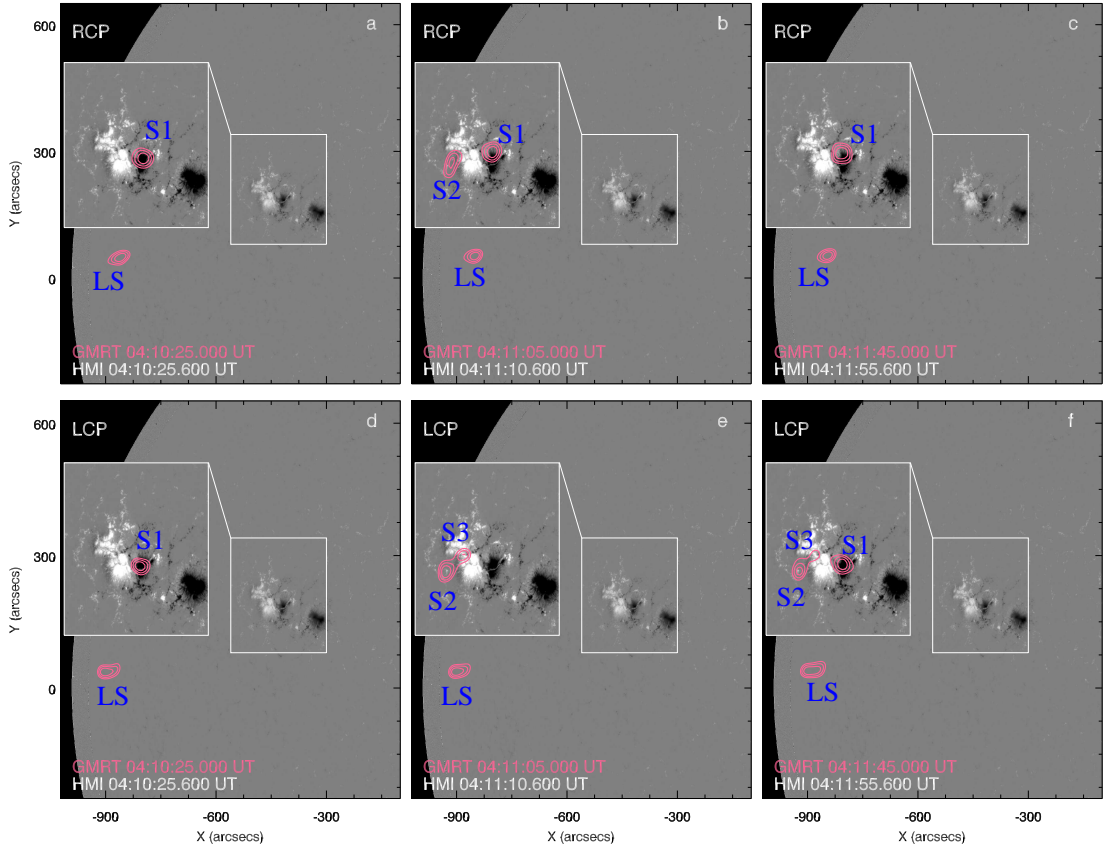

**Figure 7.** A sequence of *SDO/HMI* images overlaid by *GMRT* 610 MHz source contours (pink) at right circular polarisation (RCP) (top row) and left circular polarisation (LCP) (bottom row). The columns, from left to right, show *GMRT* contour images at the onset of the 610 MHz flare at 04:10:25 UT, at the 610 MHz flare peak at 04:11:05 UT, and after the 610 MHz flare peak at 04:11:45 UT, respectively. Source LS is seen throughout in both RCP and LCP images, while source S1 is seen in both RCP and LCP images during pre- and post-burst periods, and only in LCP images during the burst. Sources S2 and S3 have first appeared during the burst with S2 seen in both RCP and LCP images, while S3 seen only in RCP images.

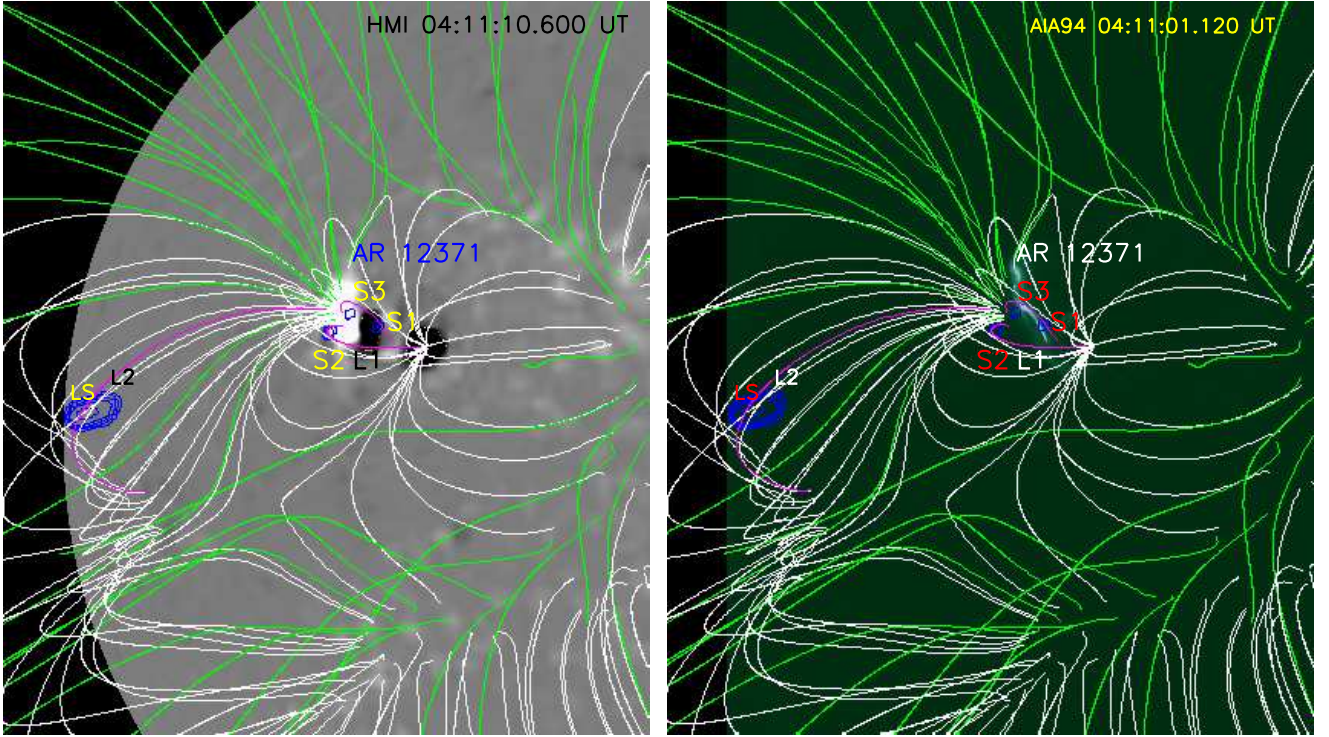

**Figure 8.** Shows PFSS computed coronal magnetic field lines of closed loops (white) and open fields (green) overlaid on background images of *SDO/HMI* (left panel) and *SDO/AIA* 94 Å (right panel). Also, overplotted in blue are contours of sources LS, S1, S2 and S3 imaged at *GMRT* 610 MHz. The magnetic loops associated with *GMRT* sources S2 and LS are shown by the solid pink curves, labelled as L1 and L2, respectively.

Using the PFSS extrapolations and a coronal density model (see Appendix A), we estimate the heights of the sources S2 and LS, which are, respectively, $\sim 58$ Mm or 1.08 $R_s$ and $\sim 127$ Mm or 1.18 $R_s$. While the sources S1 and S3 are found at the height of $\sim 23$ Mm or 1.02 $R_s$ (see Appendix A). From the PFSS extrapolation, we estimated the coronal magnetic field strengths at the locations of *GMRT* sources S1, S2, S3, and LS, which are found to be about 72 G, 40 G, 127 G and 1.3 G, respectively. Since we obtained the heights and magnetic field strengths at the location of *GMRT* sources, using the density values from the coronal density model and the PFSS extrapolations, we computed the ratio of fundamental plasma frequency ($f_p = 8980 \sqrt{n_e}$) to gyro-frequency ($f_b = \frac{eB}{2\pi m_e}$) and found a value $> 1$ for all the *GMRT* sources.

4.6. *Metric dynamic spectra*

The decimetric dynamic spectra (figure not shown here), from the *SBRS/YNAO* in the frequency range 625–1500 MHz and from the *Miangtu Spectra Radio Heliograph* (*MUSER*) (Yan et al. 2012) in the frequency range 400–2000 MHz, were examined. We also checked the radio dynamic spectra from the publicly available e-callisto network of solar radio spectrometers database. There appeared to be no evidence of type III bursts in the decimetric frequency range, covering the frequency of 610 MHz, during the radio flare maximum. However, metric type III bursts were identified in the dynamic spectra of *SBRS/YNAO* during this period. Fig.9a shows *YNAO* metric spectra, from 04:09:54–04:11:42 UT, depicting the occur-rence of type III bursts. The upper and lower panels of Fig.9a, respectively, show the spectra of LCP and RCP. Many type III bursts with different start frequencies and bandwidths are observable from Fig.9a. One such type III burst occurring at 04:10:23 UT is indicated by white arrows in Fig.9a, which has started at a high frequency compared to the other bursts, having duration of 0.5s, bandwidth of ∼130 MHz and negative frequency drift rate of 65 MHz s$^{-1}$. Note that the *GMRT* decimetric sources S1, S2, S3 and LS showed a quasi-periodic burst peak, which started just at this time.

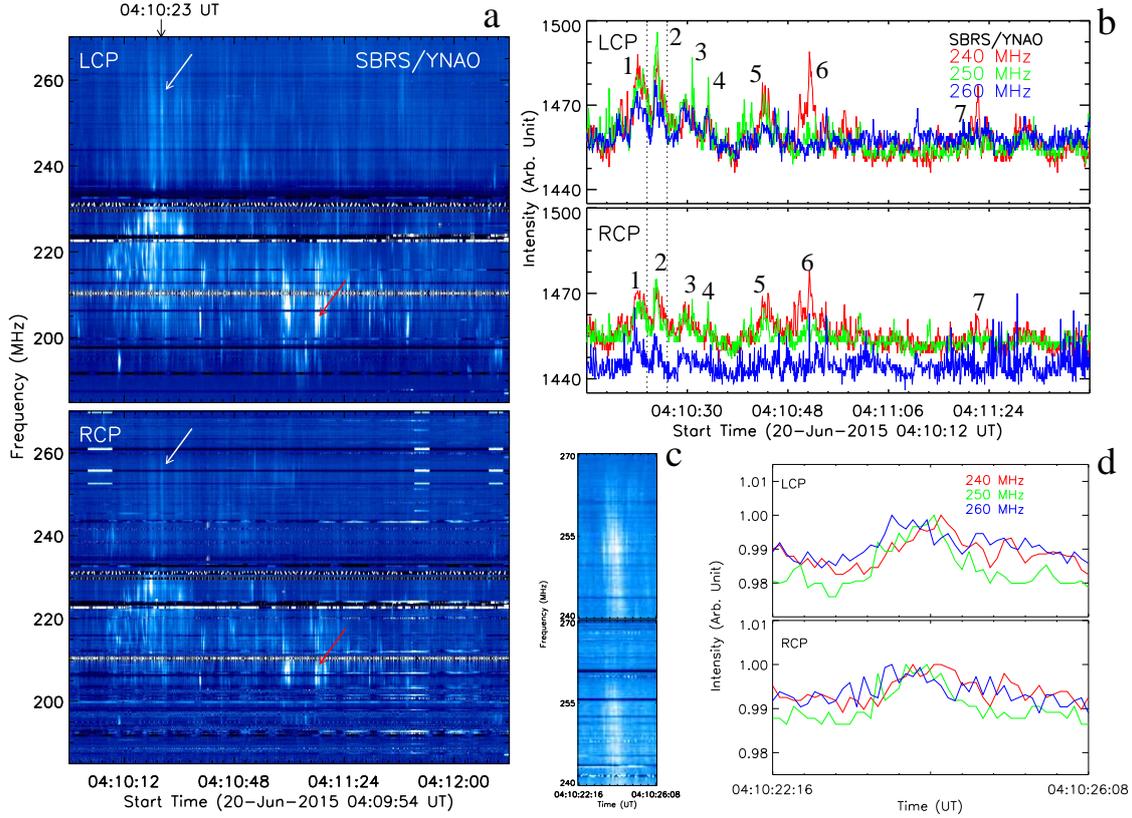

**Figure 9.** a) The enhanced view of metric radio dynamic spectra from *SBRS/YNAO*, between 04:09:54 UT and 04:11:42 UT, in LCP (top) and RCP (bottom), depicting the type III burst of interest and other narrow band type III bursts is shown. The type III burst of interest at 04:10:23 UT is marked by arrows. b) The lightcurves at selected frequencies are shown to depict the drift of bursts from higher to lower frequencies. c-d) Shown, respectively, the enhanced view of spectra and lightcurve of the type III burst of interest.

For a clear view of the type III bursts, their lightcurves at the few selected frequencies are depicted in Fig.9b. A few type III bursts with high start frequencies are indicated by numbers in Fig.9b. Also, bright narrow band metric burst emissions have been noticed in the frequency range 200–220 MHz, between 04:10:48 and 04:11:24 UT corresponding to the 610 MHz radio flare peak, from both LCP and RCP spectra as shown in Fig.9a and are indicated by red arrows. The spectra and the lightcurves of the type III burst of interest occurring at 04:10:23 UT, indicated by the white arrows in Fig.9a and demarcated between the vertical dotted lines in Fig.9b, have been shown at the few selected clean frequency channels in Fig.9c,d, depicting the negative frequency drift of the burst.

The type III bursts are noticed in both LCP and RCP metric spectra (Fig.9a), however, they show strong emission in the LCP spectra. This is also the case for our type III burst of interest, that shows a strong emission in LCP (Fig.9d). Thus, the metric bursts are weakly left circular polarised. The frequency drift rate of the type III burst noticed at 04:10:23 UT was used (see Appendix B) to estimate the electron beam velocity, which we found to be ∼0.13c. The beam velocity of ∼0.13c corresponds to the kinetic energy of ∼29 keV.

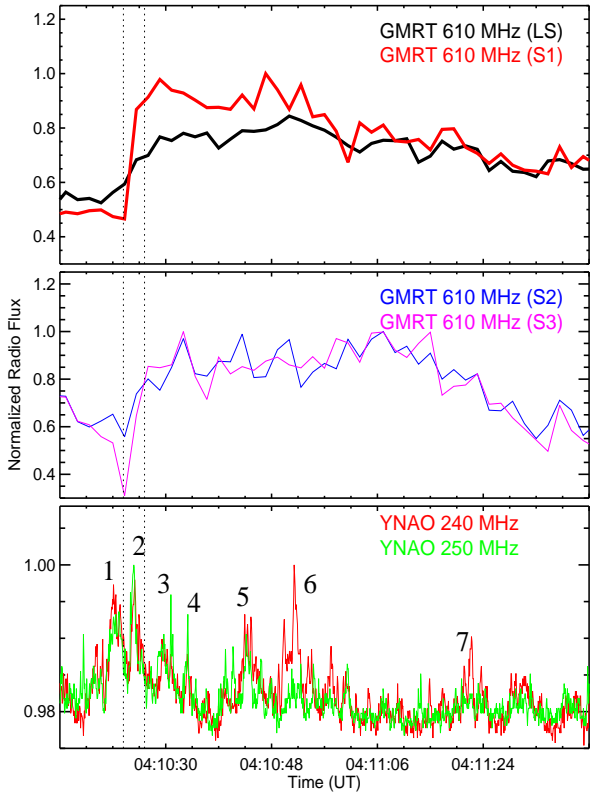

**Figure 10.** Shows the normalised radio fluxes of *GMRT* sources LS and S1 (top panel), *GMRT* sources S2 and S3 (middle panel), and type III bursts at the selected frequencies of *SBRS/YNAO* (bottom). Several type III bursts, numbered from 1 to 7, are noticed in the metric frequencies of *SBRS/YNAO* associated with the rise of the *GMRT* 610 MHz burst activity. Also, indicated between the vertical dotted lines is a strong type III burst, which is noticed around 04:10:23 UT, just at the start of rising activity of *GMRT* sources.

The temporal association of the *SBRS/YNAO* metric type III bursts and the *GMRT* 610 MHz burst activity is shown in Figure 10, which shows, in the top panel, the temporal profiles of *GMRT* sources LS and S1, from 04:10:12 - 04:11:42 UT during the 610 MHz flare maximum. The middle panel shows the temporal profiles of *GMRT* sources S2 and S3, and the bottom panel shows the temporal profiles of the type III bursts at the selected frequencies of the *SBRS/YNAO* metric spectra. Several type III bursts are clearly observable in association with the rise of the *GMRT* burst activity. In fact, a strong type III burst is noticed just at the start of the *GMRT* burst activity at 04:10:23 UT, indicated by the vertical dotted lines.

## 5. DISCUSSION

We have presented, in this paper, the first high time cadence (0.5s) solar radio maps imaged by *GMRT* at the frequency of 610 MHz during two (a C1.4 and a M1.0) flares and a CME. The flares and CME took place on 20 June 2015 in an active region AR12371. It is seen from the high spatial resolution and dynamic range images of *GMRT* 610 MHz that the radio sources, during the M1.0 flare and the CME onset, are located just above the flaring active region. In contrast, the radio source, namely LS, during the C1.4 flare, is located of about 500″ from the flaring active region. This is unique and for the first time, to the best of our knowledge, a radio burst source during a flare emission is reported in the decimetric wavelength range, that is located so far from the flaring site.

As mentioned earlier, the location of radio sources away from the flaring active region has been commonly observed in the metric frequencies. This can be understood from the fact that the radio sources at the low frequencies are usually located higher in the corona (Pick & Vilmer 2008). However, in the present case, the frequency of observation is in the decimetric range, but still the radio source LS is located even farther than the reported cases in the metric frequency range. The question of interest, there-fore, how is that possible? Weak radio burst sources *viz.* S1, S2, and S3, are also identified during the radio flare maximum of the C1.4 flare, which are, unlike the radio source LS, located near the flaring active region.

Based on our multi-wavelength analysis and PFSS magnetic field extrapolation, in the following, we discuss the question of interest, the location of electron acceleration region and the nature of the *GMRT* radio sources.

### 5.1. *The electron acceleration region*

From our study, we have shown that all the *GMRT* radio sources, widely separated in space, exhibit a close temporal correlation. Further, from our PFSS extrapolation, a common origin of the magnetic field lines intersecting the *GMRT* radio sources has been observed. All the field lines has originated in the positive polarity region P1 of the flaring active region. It led us to suggest that the source electrons or the exciters of the *GMRT* radio sources were originated from a common electron acceleration site, which is likely located near the flaring active region. It is known that the interaction sites of the small flaring loops and the larger loops could serve as the preferred sites for the electron acceleration and the energy release (Pick & Vilmer 2008). So the energetic electrons could originate from a common interaction site of the low lying flare loops and the high arching loops. The accelerated electrons from this site were injected simultaneously into different magnetic field lines, which subsequently propagated along the field lines and emitted at the respective locations of the *GMRT* radio sources.

Also, the observed temporal association of the metric type III bursts and the *GMRT* radio burst activity apparently indicates that they can be causally related. This suggests that the exciters of the radiating electrons, resulting the metric type III bursts and the *GMRT* burst activity, were originated from the same electron acceleration site. In other words, the open field lines in which the type III excited electrons propagate and the closed loops in which the exciting agent of *GMRT* radiating electrons travel could be interacting near a common interaction site. From our PFSS extrapolations, we have seen the divergent open field lines located above the positive polarity region P1 of the flaring active region. It is known that the open field lines usually appear divergent (Aschwanden et al. 1992) due to the propagation of accelerated electrons along them. This suggests that the electrons may originate from an acceleration site located in the lower corona. It further supports that the open field lines and the closed loops could be interacting at a common site, located near the flaring active region, where the radiating electrons responsible for the bursts originated.

### 5.2. *The emission mechanism*

All the *GMRT* radio sources show high brightness temperatures, $> 10^{11}$ K, suggesting that they are generated by a coherent emission mechanism, that is – a plasma or an ECM radiation process. For ECM emission to be possible, $f_p$ must be less than $f_b$ at the emission site. Instead, based on a coronal density model and PFSS magnetic field extrapolations, we have found a ratio $f_p$ to $f_b > 1$ for all the *GMRT* sources at their respective observed locations. This suggests that the *GMRT* sources could be due to a plasma emission process.

Also, from GMRT polarisation maps, we have shown that sources S1 and S3 are probably due to fundamental radiation, while sources S2 and LS are likely to be harmonic radiation. The estimated density scale heights for a frequency of 610 MHz for fundamental and harmonic radiations, as obtained using a coronal electron density model by Aschwanden & Benz (1995), would be 23 and 41 Mm, respectively. Thus, the density scale heights for sources S1 and S3 would be 23 Mm, while that for sources S2 and LS would be 41 Mm. However, the ac-tual height estimates for sources S2 (∼58 Mm) and LS (∼127 Mm), obtained from comparing our images and PFSS extrapolations (see Appendix A), are different from their density scale height of 41 Mm. Hence, one needs to normalise the Aschwanden density model by a suitable multiplication factor so as to reconcile the density scale height with the height obtained from the PFSS extrapolation. It is found that a 2-fold Aschwanden electron density model could explain the estimated height for the harmonic source S2. Thus, the normalized estimated heights, using the 2-fold Aschwanden electron density model for the fundamental sources S1 and S3 would be ∼28 Mm.

Even if we assume the source LS is a second harmonic emission source, the 2-fold Aschwanden density model, however, couldn't explain its estimated height from the PFSS extrapolation. This suggests that if the source LS is due to plasma radiation, it should be generated lower in the corona. The plasma radiation, however, could have escaped into the higher corona by a wave-ducting mechanism along a density depleted tube. From our study, we have shown that the source LS is located on the top of a high arching loop with one of its ends anchored near the flaring active region. The high arching loop could have acted as a density depleted duct. Also, as mentioned earlier, a common electron acceleration site is located near the flaring active region. This suggests that the flare-associated upward streaming electrons inside the tube could have produced the source LS by a plasma emission process, somewhere near the flaring active region. If this is the case, then the requirement of ECM condition $f_p < f_b$ (Morosan et al. 2016) may also be satisfied considering the much stronger magnetic field in the core region of the AR. Thus, ECM may also be a viable mechanism for the source LS.

A possible ECM process was proposed by Wang (2015), who suggested that flare-associated streaming electrons, while propagating in a density depleted tube, can be pitch-angle scattered by enhanced turbulent Alfvén waves. As a result, the electrons exhibit a velocity distribution, which is unstable to ECM instability. It is likely that the flare-associated upward streaming electrons, in this case, after pitch-angle scattered by the turbulent Alfvén waves, could have resulted in an ECM instability. The instability, in turn, would have generated o- or x-mode electromagnetic waves near the second harmonic of the gyro-frequency, that is, at ∼610 MHz. Hence, the source LS could be generated near the flaring site either by a plasma or an ECM radiation process. However, it could have escaped into the higher corona due to the wave ducting effect.

On the other hand, unlike the source LS which has been identified throughout the radio flare, the sources S1, S2 and S3 have been identified at the radio flare maximum. As estimated in our study from a *RHESSI* spectral analysis, we have found that the temperature of the ambient plasma during that period has been high, that is, 8–9 MK. Hence, the free-free absorption would be low allowing the detection of radio sources S1, S2 and S3. One can argue the possibility of ECM emission for the sources S1, S2, and S3, so we re-computed the ratio $f_p$ to $f_b$, based on their normalized heights using the 2-fold Aschwanden electron density model, at their respective observed locations. But we found a ratio > 1. This suggests that the *GMRT* sources S1, S2 and S3 were likely due to a plasma emission process.

### 5.3. *The wave mode*

It is known that fundamental plasma emission is highly circular polarised in the sense of the o-mode as the x-mode can not escape above a cut-off frequency of $f_p + f_b/2$ (Kundu et al. 2006; Nindos et al. 2008). Harmonic plasma emission is though weakly polarised, however, it is often polarised in the sense of o-mode (Nindos et al.

2008). It is found that, except for the source LS, the *GMRT* sources S1, S2 and S3 are by a plasma emission process, hence, they could be polarised in the sense of o-mode. From our study, we have shown that sources S1 and S3 are, respectively, strongly RCP and LCP, and the directions of the magnetic fields, from the PFSS extrapolation, in sources S1 and S3 are, respectively, downwards and upwards directed. Thus, the dominant wave modes of source S1 and S3 would be o-mode. Source S2 is, on the other hand, weakly LCP and located on the loop with one of its end anchored in regions of upward directed magnetic field lines. Thus, the dominant wave mode of source S2 would be o-mode. The metric type III bursts reported in the paper are also weakly LCP bursts. Thus, the dominant wave mode of the type III bursts would be o-mode if their burst sources are also located in the regions of the upward directed field lines.

The source LS is weakly LCP. If we consider it is due to plasma radiation then the dominant wave mode of source LS would be o-mode provided it is located in the regions of the upward directed field lines. However, if we consider the source LS is due to ECM radiation then the value of $f_b$ for the source LS at its generation site would be ∼305 MHz. It is known that the wave modes of ECM emission depend on the ratio of $f_p$ to $f_b$ at the true source location or the generation site (Melrose & Dulk 1982; Sharma & Vlahos 1984; Régnier 2015; Wang 2015). Wang (2015) also suggested that if the ratio $f_p$ to $f_b$ is in the range 0.4–1.4, then only a harmonic o- or x- mode structureless burst would be observed instead of a fundamental and harmonic pair burst. However, for the present case, the value of $f_p$ at the generation site could not be determined. Therefore, the wave mode associated with the source LS is unclear if it is due to an ECM process.

## 6. SUMMARY AND CONCLUSION

In the present work, we have presented high temporal and spatial radio images, produced using *GMRT* 610 MHz observations obtained during a C1.4 class solar flare on 20 June 2015. We have reported a strong decimetric radio source, located far from the flaring active region. Also, we have reported weak decimetric radio sources identified during the 610 MHz flare maximum. The weak radio sources are however, located near the flaring site. Further, they show a close temporal correlation with the strong radio source and as well with the metric type III radio bursts identified in the *SBRS/YNAO* metric dynamic spectra.

Based on our investigation of a multi-wavelength analysis and PFSS extrapolations, we have suggested that the source electrons of decimetric radio sources and metric type III bursts were originated from a common electron acceleration site. The acceleration site is likely located near the flaring active region. In addition, we have argued that the observed weak radio sources are o-mode waves produced by a plasma emission process. The strong decimetric source, on the other hand, can be generated either by a plasma or an ECM emission process. Its location far from the flaring site is presumably caused by the wave ducting of the emitted coherent radio waves that escaped along a density-depleted high arching loop.

Although the single frequency *GMRT* 610 MHz observations provide important insights on the flare-accelerated electrons. However, they are unable to track the streaming electrons associated with the electron acceleration process, because they traverse a wide range of coronal heights as they propagate out from the electron acceleration site. We expect future high cadence (25 to 200 ms) and high spatial (1.3″ to 50″) resolution, and multi-frequency (0.4–2 GHz) imaging observations from the *MUSER* can help in studying the details of electron beam trajectories and can provide useful insights into the source regions of electron acceleration at decimetric frequencies.

The authors thank the staff of the *GMRT* for the help during the observations. The authors also acknowledges *SDO/AIA*, *SDO/HMI*, *NASA/RHESSI*, and *GOES* teams making the data publicly available. SKB is supported by PIFI (Project Number: 2015PM066) program of Chinese Academy of Sciences and by NSFC (Grant no. 117550110422). LJC and GG acknowledge supports by NSFC (Grant no. 11573043 and Grant no. 11403099, respectively). SKB acknowledges the scientific inputs from P. Subramanian, Baolin Tan and Jie Huang. In addition, the authors thank the reviewer for his constructive comments and suggestions which significantly improved the paper.

## APPENDIX

### A. ESTIMATION OF HEIGHTS OF GMRT RADIO SOURCES

The comparison of spatial locations of *GMRT* radio sources with PFSS computed coronal magnetic fields showed that they were loop top radiations (see Fig.8). In order to estimate the radial heights of *GMRT* radio sources, we identified the coronal loop, obtained from PFSS extrapolation, on which the centroids of *GMRT* radio source was actually located and also the apex of loop was situated close to the *GMRT* radio source. Fig.11 shows the PFSS computed coronal magnetic field lines overlaid on the background *HMI* magnetogram at 04:11:05 UT. The centroids of all *GMRT* radio sources (S1, S2, S3 and LS) are overplotted on the magnetogram as star symbols. It is seen from Fig.11 that the centroids of sources S2 and LS are located on the coronal loops, labelled as L1 and L2 and shown by the solid pink curves, whose loop apexes were close to the source centroids of the former. The apexes of loops L1 and L2, as estimated from PFSS extrapolation, were located at heights of ∼58 Mm and ∼140 Mm, respectively. The estimated heights are radial heights from the solar surface. Thus, the radial heights of GMRT sources S2 and LS could be ∼ 58 Mm or 1.08 $R_s$ and ∼140 Mm or 1.2 $R_s$, respectively.

It may though happen that the apex of loop could be located at a non-radial position. The actual height of radio sources then could be different, which can only be estimated by knowing the inclination of the loop apex to the line-of-sight direction from Earth. To find the inclination of loop apex, the views of the loop from two different angles such as *SDO* (Earth's view) and *STEREO* (*Solar Terrestrial Relation Observatory*) should be available. The twin spacecrafts *STEREO* A and B, during this event, were however, separated from the Sun-Earth line by angles of 177° and 172°, respectively, making it difficult to find the inclination. From the images though the apparent inclination could be estimated (Iwai et al. 2014), which is ≤ 2° for source S2, and about 25° for source LS. Thus, the actual heights of *GMRT* source S2 would be same as the radial heights of ∼ 58 Mm or 1.08 $R_s$, while the actual height of source LS would be 127 Mm or 1.18 $R_s$.

It is found that sources S1 and S3 are located in the loop legs, we thus estimated the density scale heights for sources S1 and S3 using a coronal electron density model by Aschwanden & Benz (1995). In the Aschwanden & Benz (1995) density model, the electron density in the lower corona (for h < 160 Mm) can be obtained using the following equation:

$$n_e = n1.(\frac{h}{h1})^{(-p)} \tag{A1}$$

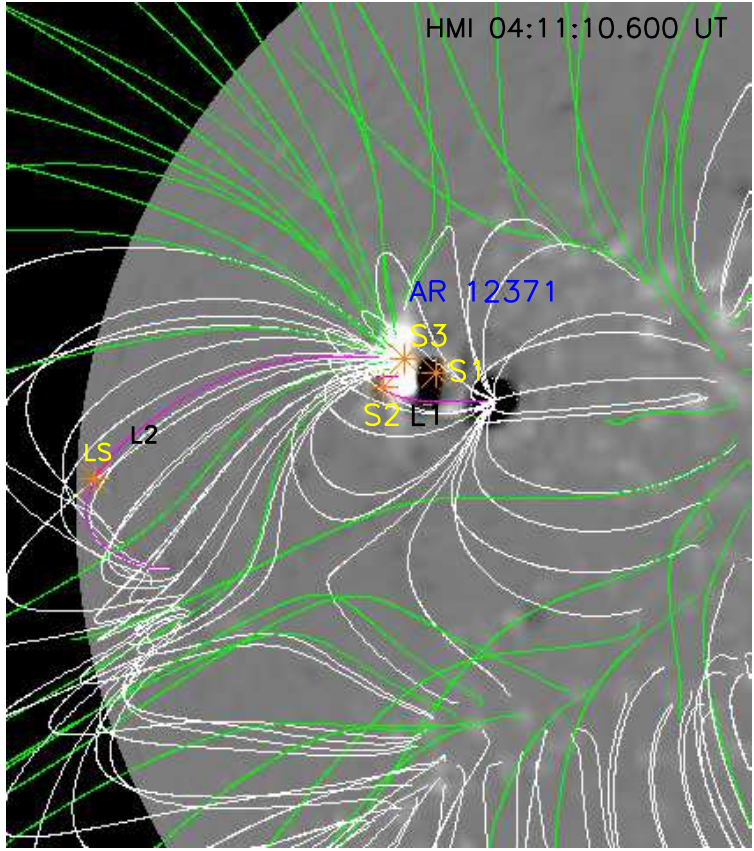

**Figure 11.** The PFSS computed coronal magnetic field lines are shown in white (for closed field lines ) and green (for open field lines), overlaid on a background *HMI* image, at the radio flare maximum. Also, overplotted as star symbols in orange are centroids of *GMRT* 610 MHz sources LS, S1, S2 and S3. The magnetic field loops associated with sources S2 and LS are shown by the solid pink curves, labelled as L1 and L2.

where $n1 = nQ.exp(-p)$, p=2.38, $\lambda = 6.9 \times 10^{10}$ cm, $h1 = 1.6 \times 10^{10}$ cm, $n_Q = 4.6 \times 10^8 cm^{-3}$ $n_e$ is known from the frequency of observation by the relation $f = s.8980 \sqrt{n_e}$ (s=1 for sources S1 and S3). Thus, the estimated density scale heights of sources S1 and S3 would be $\sim 23$ Mm or 1.02 $R_s$.

### B. ESTIMATION OF VELOCITY OF TYPE III ELECTRON BEAM

The velocity of the electron beam can be estimated from the frequency drift rate, $\frac{df}{dt}$, of the type III burst and knowing the scale height, $H_n$ at the drift rate measured frequency f:

$$v \equiv 2H_n \cdot \frac{1}{f} \cdot \frac{df}{dt} \tag{B2}$$

The $H_n$ is estimated using the St. Hilarie density model (Saint-Hilaire et al. 2013) knowing $n_e$ at the measured frequency, f, which is $\sim 2.2 \times 10^8$ cm$^{-3}$. The St. Hilarie model is chosen because it is a solar wind like density model of the form $n_e$ (r) = C $(h/R_s)^{-2}$, where h is heliocentric distance and C is a normalisation constant with a value of $1.2 \times 10^6$ cm$^{-3}$. Thus, the estimated apparent velocity of the electrons was found to be $\sim$0.16 c. To estimate the real velocity of the electron, we need the angle of propagation between the beam direction and the line of sight. However, we didn't know the angle and without knowing the angle, the minimum possible real electron velocity, $v_{min}$ could be estimated following the method described in Carley et al. (2016):

$$v_{min} = \frac{cv}{c+v} \tag{B3}$$

The real velocity of the electron beam was therefore $\sim 0.13c$.